\documentstyle[11pt,paspconf,psfig]{article}

\begin{document}

\title{
 VLBI Polarisation Images of the Gravitational Lens B0218+357
  }
\author{
 A.J. Kemball$^1$, A.R. Patnaik$^2$, R.W. Porcas$^2$
  }
\affil{
$^1$~National Radio Astronomy Observatory, Socorro, USA \\
 $^2$~Max-Planck-Institut f{\"u}r Radioastronomie, Bonn, Germany \\
  }

\begin{abstract}

  We present preliminary polarisation VLBI maps of the gravitational
  lens system B0218+357, made at 3 different frequencies.
\end{abstract}

\keywords{radio polarisation}

\section{Summary}

We have made VLBI polarisation observations of the gravitational lens system B0218+357 (Patnaik et al, 1993).
Observations at 8.4 GHz were made on 9 May 1995, 
using the NRAO VLBA together with the 100m Effelsberg telescope,
and at 22 and 43 GHz on 29 May 1996 using the VLBA alone.
Reduction of the data was made using standard procedures in
the AIPS software package.

Preliminary maps are presented in Figure 1. 
The "core" and "knot" components, seen by Patnaik et al (1995) at 15 GHz,
appear at all 3 frequencies in both A and B images. 
The core (right) is highly polarised - consistent with the high degree 
of polarisation variability shown by this source (Biggs et al, 1999).

Both the A and B image paths are known to suffer high Faraday rotation,
with a differential RM of 980$\pm$10~rad~m$^{-2}$
between the images (Patnaik et al, {\it these proceedings}).
Our observations are not suitable for a direct determination of RMs
as they are not simultaneous,
and the polarisation angle can vary on the same timescale as the
image relative delay (Biggs et al, 1999).
However, the effect of differential rotation between
the A and B image paths
is apparent in the increase with wavelength of the difference between
the core PAs of A and B.
Indeed, the parallel PAs of the A and B cores at 43 GHz (where
Faraday rotation is negligible) nicely demonstrates a
basic property of gravitational lensing - that the PA of polarisation
is unchanged by the action of the lens, even though source structural
position angles may be changed in the images.

\begin{figure}[h]
\centerline{
{\psfig{figure=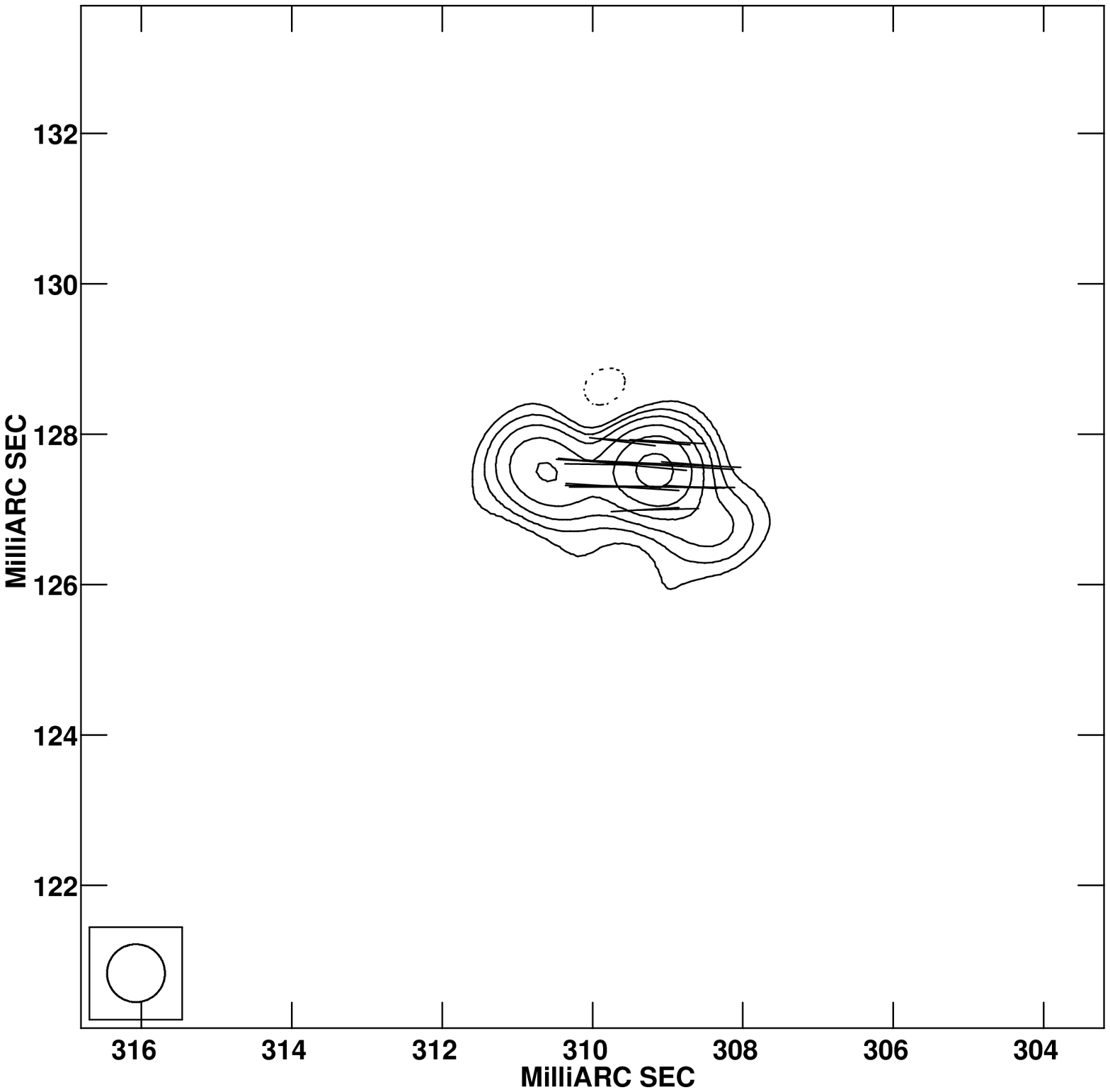,width=5.5cm,clip=}}
{\psfig{figure=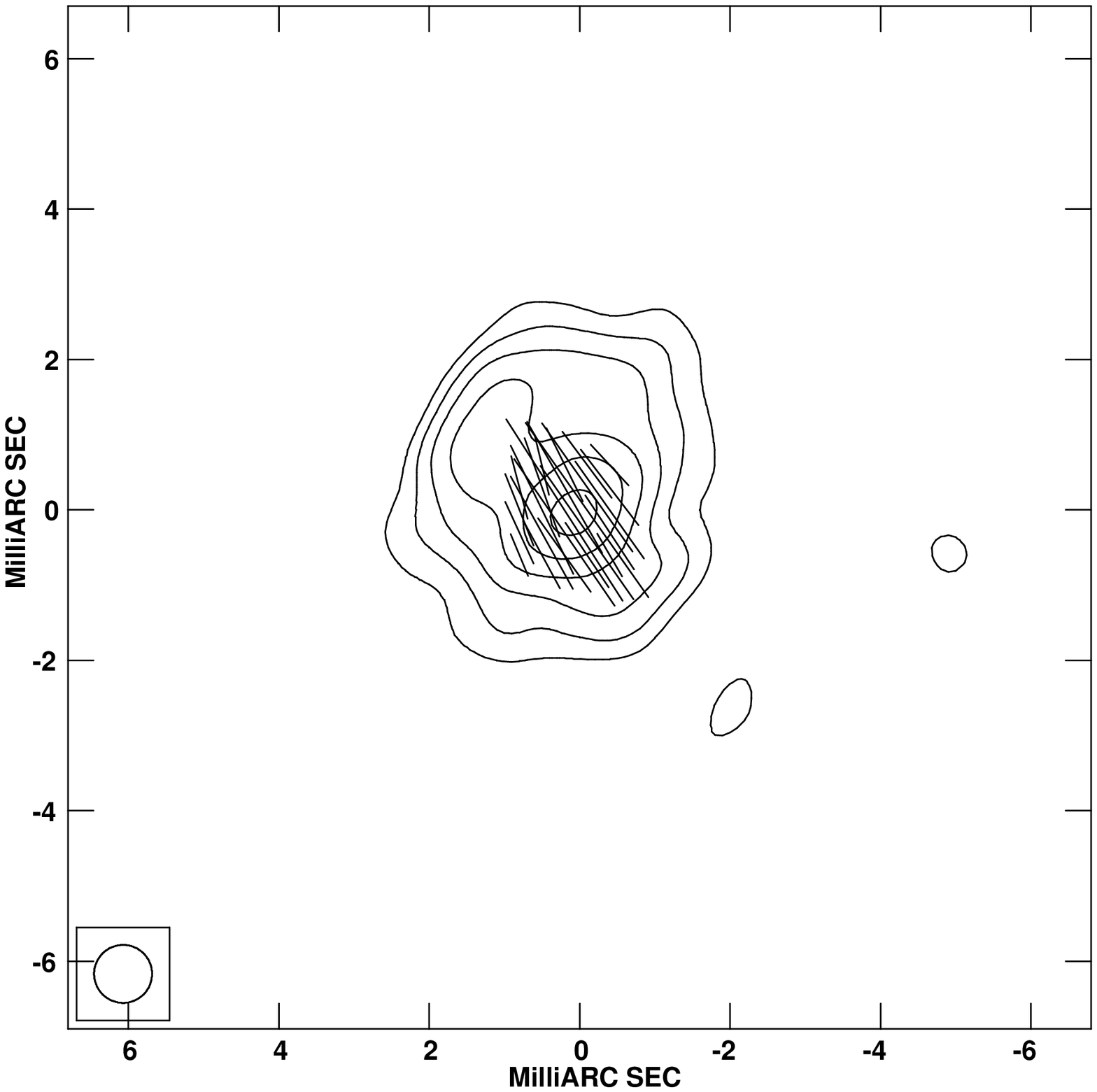,width=5.5cm,clip=}}
}
\centerline{
{\psfig{figure=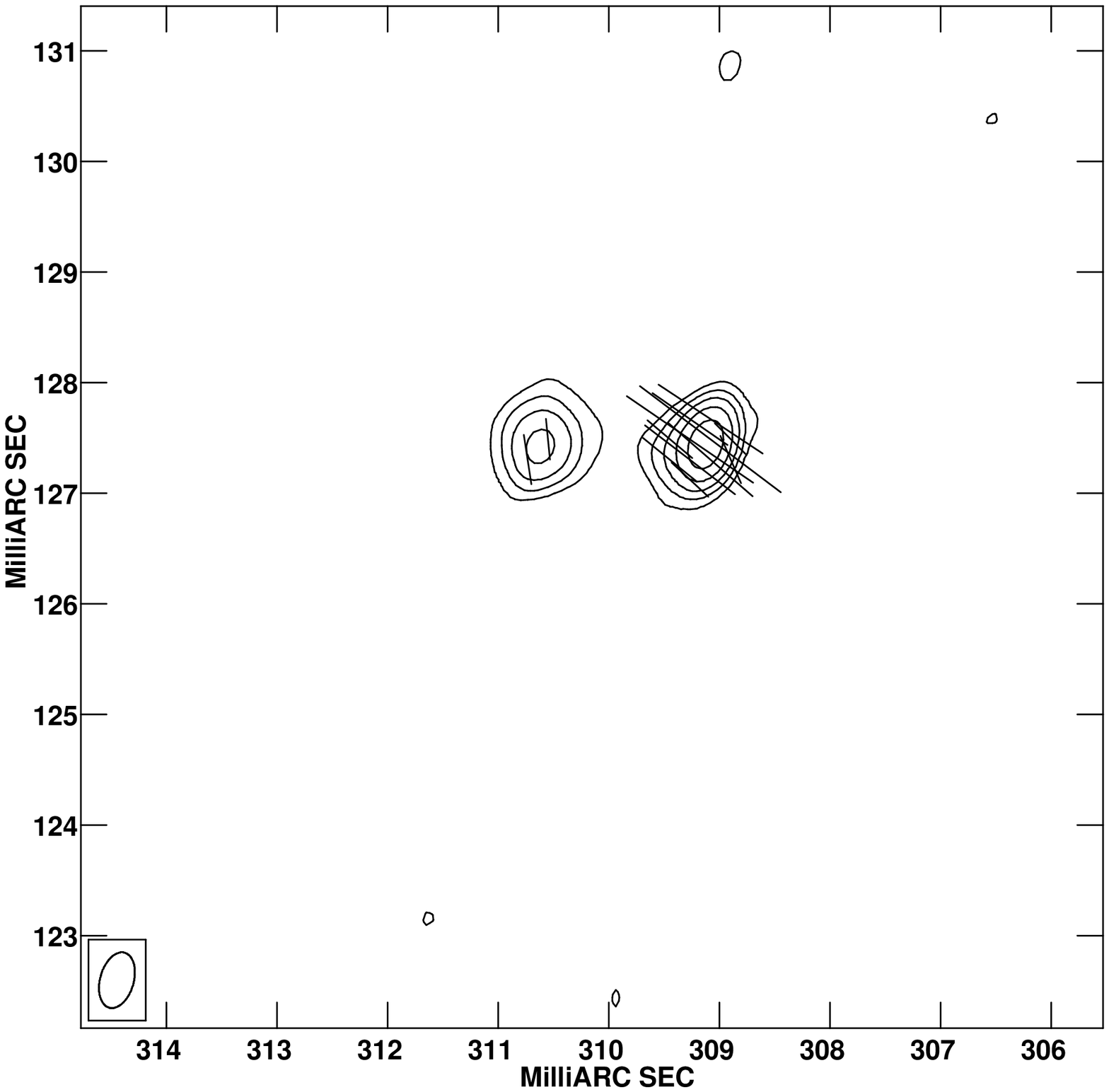,width=5.5cm,clip=}}
{\psfig{figure=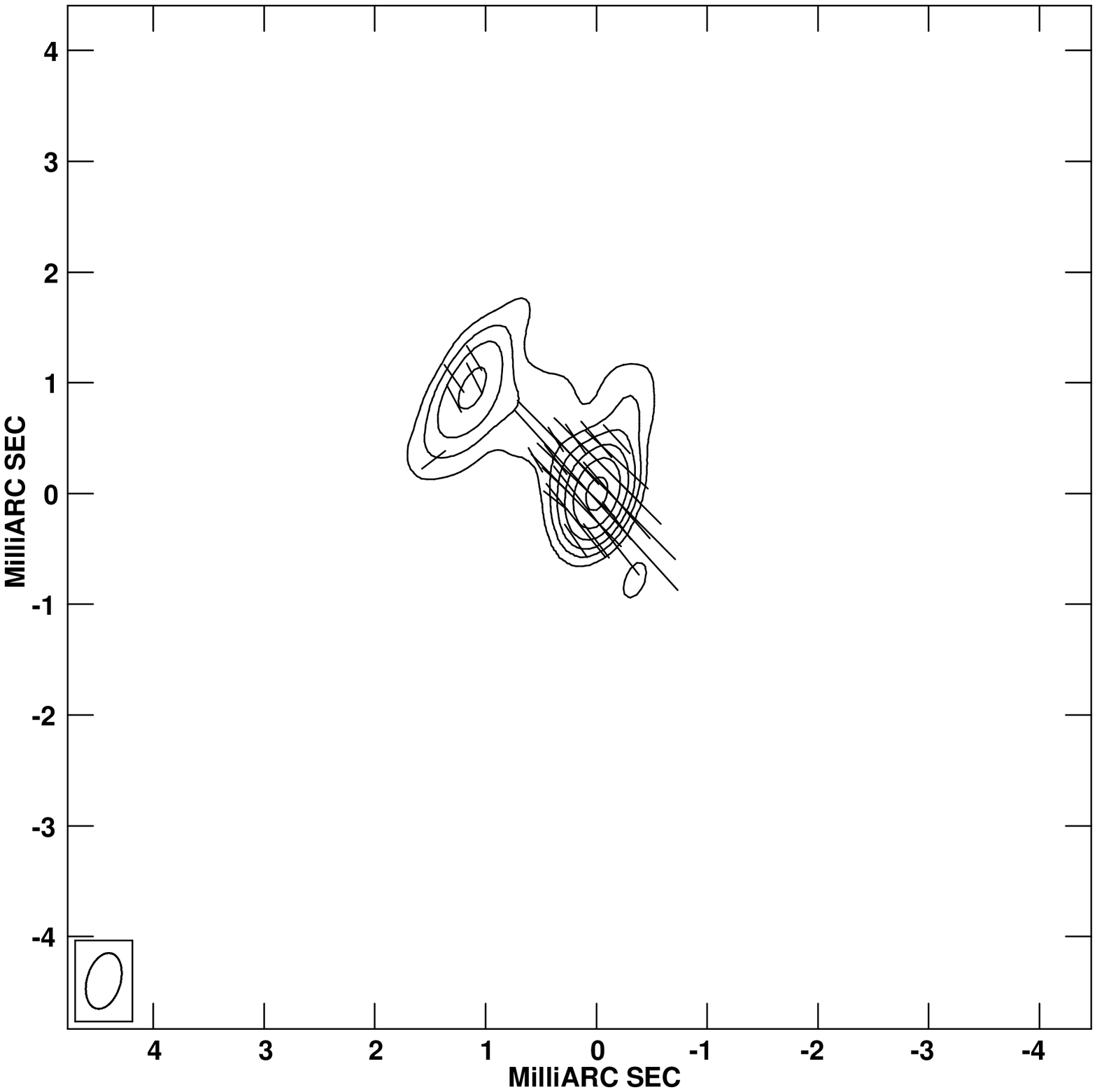,width=5.5cm,clip=}}
}
\centerline{
{\psfig{figure=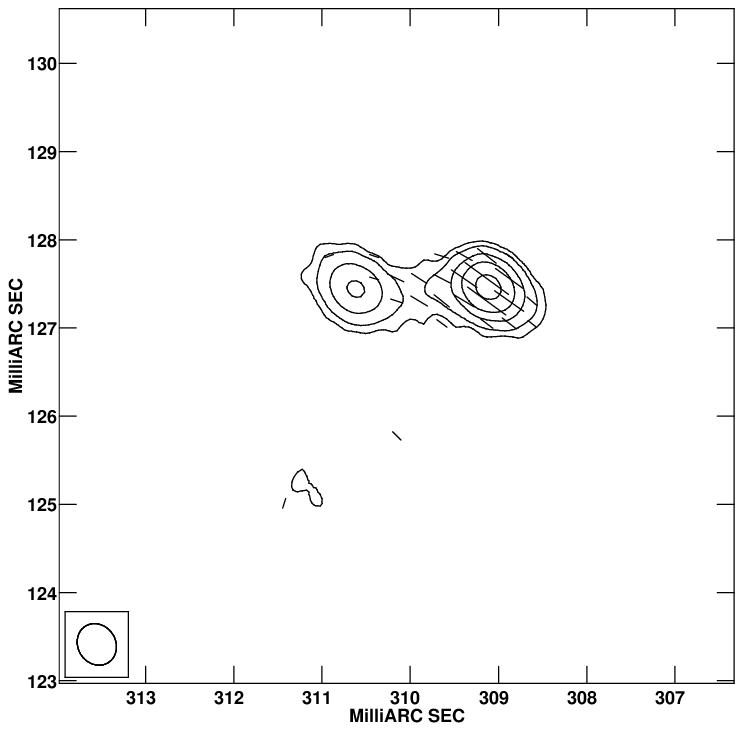,width=5.7cm,clip=}}
{\psfig{figure=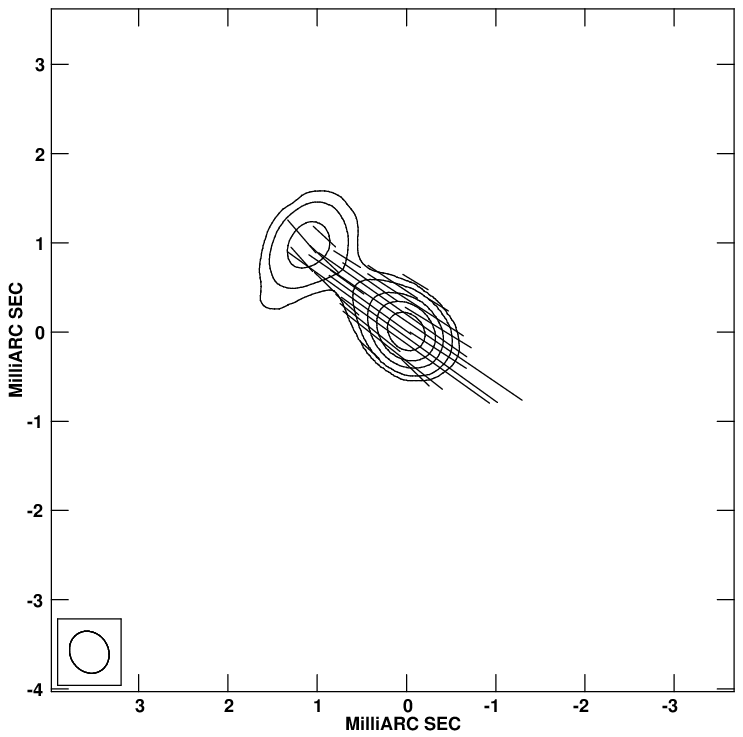,width=5.7cm,clip=}}
}
\caption{VLBI Polarisation maps of B0218+357. 
  {\bf Left:} Image B~ {\bf Right:} Image A.
  Total intensity contours
  are at relative levels: -5, 2, 5, 10, 20, 40, 80, 160; within each
  map the length of vectors are proportional to linearly polarised
  intensity and their direction is that of the radiation E field.
  CLEAN restoring beams are plotted in the lower left of each map.
  {\bf Top:}~~8.4 GHz~~{\bf Middle:}~~22 GHz~~{\bf Bottom:}~~43~GHz
  }

\end{figure}
\acknowledgments
The National Radio Astronomy Observatory is a facility
of the National Science Foundation operated under cooperative
agreement by Associated Universities, Inc.

\end{document}